# How Strings Make Do Without Supersymmetry: An Introduction to Misaligned Supersymmetry*

Keith R. Dienes[†]

*School of Natural Sciences, Institute for Advanced Study*
*Olden Lane, Princeton, NJ 08540 USA*

## Abstract

We provide a non-technical introduction to "misaligned supersymmetry", a generic phenomenon in string theory which describes how the arrangement of bosonic and fermionic states at all string energy levels conspires to preserve finite string amplitudes even in the absence of spacetime supersymmetry. Misaligned supersymmetry thus naturally constrains the degree to which spacetime supersymmetry can be broken in string theory while preserving the finiteness of string amplitudes, and explains how the requirements of modular invariance and absence of physical tachyons affect the distribution of states throughout the string spectrum.

---

# 1 Introduction: Motivation and Overview of Results

The distribution of states in string theory is an important but not particularly well-understood issue. It is well-known, for example, that string theories generically contain a variety of sectors, each contributing an infinite tower of states from the massless level to the Planck scale, and it is also a generic feature that the number of these states as a function of the worldsheet energy $n$ grows asymptotically as $e^{C\sqrt{n}}$ where $C$ is the inverse Hagedorn temperature of the theory. Beyond these gross features, however, not much is known. For example, modular invariance presumably tightly constrains the numbers of string states at all energy levels, but a precise formulation of such a constraint is still lacking. Similarly, the distribution of bosonic and fermionic string states at all energy levels is crucial in yielding the ultraviolet finiteness for which string theory is famous, yet it is not clear precisely how the actual distribution of such states level-by-level conspires to achieve this remarkable result. Of course, if the string theory in question exhibits spacetime supersymmetry, both issues are rendered somewhat trivial: there are necessarily equal numbers of bosonic and fermionic states at every energy level, the one-loop partition function vanishes, and the divergences from bosonic states are precisely cancelled by those from fermionic states. Yet how does the string spectrum manage to maintain modular invariance and finiteness in the *absence* of spacetime supersymmetry? Alternatively, to what extent can one break spacetime supersymmetry in string theory without destroying these desirable features?

Recently, there has been some progress in answering these questions, and in particular it has been shown [1] that even in the absence of spacetime supersymmetry, string spectra generically turn out to exhibit a residual cancellation, a so-called "misaligned supersymmetry". In fact, this property is completely general, and serves to describe the distribution of bosonic and fermionic states in *any* string theory which is modular-invariant and free of physical tachyons. In this paper, we shall provide a short non-technical introduction to misaligned supersymmetry, stressing only its most phenomenological features. For further details we refer the reader to Ref. [1].

# 2 Background: Three Questions

In order to gain some insight into the relevant string issues, we shall begin by providing some background into the questions raised above.

## 2.1 How does modular invariance constrain the numbers of states in string theory?

In string theory, the issues of modular invariance and the numbers of bosonic and fermionic physical states at each energy level are directly related through the one-loop partition function $Z(\tau)$. Given a torus with modular parameter $\tau$, the partition function $Z(\tau)$ is simply defined as a trace over all of the states in the theory:

$$Z(\tau) \equiv \sum_{\text{states}} (-1)^F \, e^{2\pi i \tau H} \, e^{-2\pi i \overline{\tau} \overline{H}} \, . \qquad (1)$$

Here the factor $(-1)^F$ indicates that spacetime bosonic and fermionic states contribute to $Z$ with opposite signs, and $H$ and $\overline{H}$ denote the separate Hamiltonians for the left- and right-moving worldsheet degrees of freedom. Thus, if we define the quantity $q \equiv e^{2\pi i \tau}$ and expand $Z(\tau)$ as a double power series in $q$ and $\overline{q}$,

$$Z(\tau) = \sum_{m,n} a_{mn} \, \overline{q}^m \, q^n \, , \qquad (2)$$

we see that the coefficients $a_{mn}$ of this expansion yield the net numbers of string states (bosonic minus fermionic) with right- and left-moving worldsheet energies $(m, n)$. Since the states with $m = n$ are the *physical* (or "on-shell") states which contribute to the actual spectrum of the theory, the set of state degeneracies $\{a_{nn}\}$ forms our object of interest.

Although the partition function (1) clearly resembles a statistical-mechanical partition function in which $\tau$ plays the role of an inverse temperature (or equivalently resembles a field-theoretic generating functional, with $\tau$ analogous to a Schwinger proper time), in string theory the parameter $\tau$ instead describes the geometry of the underlying torus. This is, however, a crucial distinction, for it turns out that for any value of $\tau$, each of the quantities $\{\tau, \tau+1, -1/\tau\}$ also describes the same torus. Thus for consistent string theories, we must in fact impose a constraint which has no field-theoretic or statistical-mechanical counterpart:

$$Z(\tau) = Z(\tau+1) = Z(-1/\tau) \, . \qquad (3)$$

This constraint is known as *modular invariance*. While it is clear that the constraint (3) profoundly restricts the state degeneracies $\{a_{nn}\}$ which can appear in Eq. (2), it proves surprisingly difficult to turn the constraint (3) into a constraint on the actual degeneracies $\{a_{nn}\}$. Indeed, the generic behavior of $\{a_{nn}\}$ required by modular invariance is almost completely unknown. "Misaligned supersymmetry" will turn out to provide such a constraint.

## 2.2 How does the presence of unphysical tachyons affect the balance between bosonic and fermionic states in string theory?

A slightly more physical way of addressing the same issue is to focus instead on the tachyonic states which generically appear in string theory. Since the worldsheet energies $(m, n)$ of a given string state correspond to its (mass)$^2$ in spacetime, string states with negative $m$ or $n$ correspond to spacetime tachyons. Now, it is well-known that any string theory in which there are no tachyons must necessarily have equal numbers of bosonic and fermionic states at all mass levels. This is ultimately a consequence of modular invariance, which in this simple case can be used to relate the numbers of very low energy states such as tachyons to the numbers of states at higher mass levels. However, while the requirement that there be no *physical* tachyons is necessary for the consistency of the string in spacetime, *unphysical* tachyons (*i.e.*, tachyonic states for which $m \neq n$) cause no spacetime inconsistencies and are in fact unavoidable in the vast majority of string theories (such as all non-supersymmetric heterotic strings). This is therefore the more general case. The question then arises: how do the bosonic and fermionic states effectively redistribute themselves at all energy levels in order to account for these unphysical tachyons? To what extent is the delicate boson/fermion balance destroyed?

## 2.3 To what extent can one break spacetime supersymmetry without destroying the finiteness of string theory?

A third way of asking essentially the same question is within the framework of string finiteness and supersymmetry-breaking. If we start with a string theory containing an unbroken spacetime supersymmetry, then there are an equal number of bosonic and fermionic states at each mass level in the theory (*i.e.*, $a_{mn} = 0$ for all $m$ and $n$), and consequently we find $Z = 0$. This is of course trivially modular-invariant, and the fact that such theories have $a_{nn} = 0$ for all $n < 0$ indicates that they also contain no physical tachyons. These two conditions, however, are precisely those that enable us to avoid certain ultraviolet and infrared divergences in string loop amplitudes: modular invariance eliminates the ultraviolet divergence that would have appeared as $\tau \to 0$, and the absence of physical tachyons ensures that there is no infrared divergence as $\tau \to i\infty$. For example, the one-loop vacuum energy (cosmological constant) $\Lambda$ would ordinarily diverge in field theory, but turns out to be *finite* in any modular-invariant, tachyon-free string theory. Indeed, these finiteness properties of string loop amplitudes are some of the most remarkable and attractive features of string theory relative to ordinary point-particle field theory.

If the spacetime supersymmetry is broken, however, the partition function $Z$ will no longer vanish, and bosonic states will no longer exactly cancel against fermionic states level-by-level in the theory. However, we would still like to retain the finiteness properties of string amplitudes that arise in the supersymmetric theory. What residual cancellation, therefore, must nevertheless survive the supersymmetry-breaking process? What weaker cancellation preserves the modular invariance and tachyon-free properties which are necessary for finiteness and string consistency?

## 3  Misaligned Supersymmetry: The Basic Ideas

It turns out that misaligned supersymmetry may provide an answer to all of these questions: it yields a constraint on the allowed numbers of string states which arises from modular invariance; it describes the perturbation of the boson/fermion balance due to the presence of unphysical tachyons; and it serves as the residual cancellation which is necessary for string finiteness. Indeed, it furnishes us with a constraint on those supersymmetry-breaking scenarios which maintain string finiteness, essentially restricting us to only those scenarios in which a *misaligned* supersymmetry survives. In this section we shall briefly describe the basic features of this misaligned supersymmetry, leaving the details to be found in Ref. [1].

The basic idea behind misaligned supersymmetry is quite simple. As we have said, ordinary supersymmetry may characterized by a complete cancellation of the net physical state degeneracies $a_{nn}$ for all $n$, and this in turn implies that there are equal numbers of bosons and fermions at all mass levels in the theory. In the more general case of *misaligned* supersymmetry, each of these features is changed somewhat. First, the object which experiences a cancellation is no longer the actual net state degeneracies $a_{nn}$, but rather a new object called the "sector-averaged" state degeneracies and denoted $\langle a_{nn} \rangle$. This is will be defined below. Second, just as the cancellation of the actual net degeneracies $a_{nn}$ implied equal numbers of bosonic and fermionic states at every energy level in the theory, the cancellation of the sector-averaged number of states $\langle a_{nn} \rangle$ will instead turn out to imply a subtle boson/fermion *oscillation* in which, for example, any surplus of bosonic states at any mass level of the theory implies the existence of a larger surplus of fermionic states at the next higher level, which in turn implies an even larger surplus of bosonic states at an even higher level, *etc.* Such an oscillation is quite dramatic and highly constrained, and its precise form will be discussed below.

## 3.1 The "Sector-Averaged" Number of States $\langle a_{nn} \rangle$

We begin by describing the notion of the "sector-averaged" number of states $\langle a_{nn} \rangle$. In order to do this, let us first recall how states are typically arranged in string theory.

As mentioned in the Introduction, the generic string spectrum consists of a collection of infinite towers of states: each tower corresponds to a different *sector* of the underlying string worldsheet theory, and consists of a ground state with a certain vacuum energy $H_i$ and infinitely many higher excited states with energies $n = H_i + \ell$ where $\ell \in \mathbb{Z}$. The crucial observation, however, is that the different sectors in the theory will in general be *misaligned* relative to each other, and start out with different vacuum energies $H_i$ (modulo 1). For example, while one sector may contain states with integer energies $n$, another sector may contain states with $n \in \mathbb{Z} + 1/2$, and another contain states with $n \in \mathbb{Z} + 1/4$. Thus each sector essentially contributes a separate set of states to the total string spectrum, and we can denote the net degeneracies from the $i^{\rm th}$ individual sector as $\{a_{nn}^{(i)}\}$, where $n \in \mathbb{Z} + H_i$.

For each sector $i$, let us now take the next step and imagine analytically continuing the set of numbers $\{a_{nn}^{(i)}\}$ to form a smooth function $\Phi^{(i)}(n)$ which not only reproduces $\{a_{nn}^{(i)}\}$ for the appropriate values $n \in \mathbb{Z} + H_i$, but which is continuous as a function of $n$. Indeed, these functions $\Phi^{(i)}(n)$ clearly must not only exhibit the leading exponential dependence $e^{C\sqrt{n}}$ which typifies the well-known Hagedorn behavior of the physical-state degeneracies in string theory, but must also contain all of the subleading behavior as well so that exact results can be obtained for the relevant values of $n$. It turns out that such continuations are essentially unique and relatively straightforward, and indeed there exist well-defined procedures for generating these functions [2].

Given that such functions $\Phi^{(i)}(n)$ exist, the "sector-averaged" number of states is then defined quite simply as a sum of these functions over all sectors in the theory:

$$\langle a_{nn} \rangle \equiv \sum_i \Phi^{(i)}(n) . \qquad (1)$$

Note that $\langle a_{nn} \rangle$ therefore differs quite strongly from any of the actual physical-state degeneracies $a_{nn}^{(i)}$ which arise from a given sector, and differs as well from the total physical-state degeneracies $a_{nn}$ which appear in Eq. (2). Instead, $\langle a_{nn} \rangle$ is a continuous function which represents their "average" as defined in Eq. (1).

Given this definition, then our main result is that while spacetime supersymmetry may be broken in a given string theory — implying that some of the $a_{nn}$ are necessarily non-zero — their sector-average $\langle a_{nn} \rangle$ must nevertheless vanish.* In particular, all of the exponential growth of the individual functions $\Phi^{(i)}(n)$ must somehow cancel in the sum (1). This, then, is the residual cancellation which governs the generic string spectrum, required by modular invariance and necessary for string finiteness.

### 3.2 Misaligned Supersymmetry and Boson/Fermion Oscillations

This cancellation has far-reaching implications, and in particular implies a corresponding "misaligned supersymmetry" with boson/fermion oscillations. We can easily see how this emerges by imagining a simple example, a toy string model containing only two sectors $A$ and $B$. For the sake of concreteness, let us assume that these two sectors have different vacuum energies, with $H_A = 0$ (modulo 1) and $H_B = 1/2$ (modulo 1). We thus have two separate towers of states in this theory, with degeneracies $\{a_{nn}^{(A)}\}$ situated at energy levels $n \in \mathbb{Z}$, and degeneracies $\{a_{nn}^{(B)}\}$ situated at energy levels $n \in \mathbb{Z} + 1/2$. Then if the corresponding functional forms that describe these degeneracies are $\Phi^{(A)}(n)$ and $\Phi^{(B)}(n)$ respectively, then the cancellation of the sector-averaged number of states $\langle a_{nn} \rangle$ for this theory implies that

$$\Phi^{(A)}(n) + \Phi^{(B)}(n) = 0 . \tag{2}$$

It is important to realize that this result does not imply any direct cancellation between bosonic and fermionic states in this theory, for Eq. (2) represents merely a cancellation of the *functional forms* $\Phi^{(A,B)}(n)$. Indeed, despite the result (2), the total physical-state degeneracies $\{a_{nn}\}$ for this theory do not vanish for any particular $n$. Rather, due to the misalignment between the two sectors in this hypothetical example, the actual value of $a_{nn}$ will be $\Phi^{(A)}(n)$ if $n \in \mathbb{Z}$, or $\Phi^{(B)}(n)$ if $n \in \mathbb{Z} + 1/2$. Indeed, there exists no single value of the energy $n$ for which the actual physical-state degeneracy $a_{nn}$ is described by the vanishing sum (2).

Perhaps even more interestingly, this result implies that we cannot even *pair* the states situated at corresponding levels in the $A$ and $B$ sectors, for while the net number of states at the $\ell^{\rm th}$ level of sector $A$ is given by $\Phi^{(A)}(\ell)$, the net number of states at the $\ell^{\rm th}$ level of sector $B$ is given by $\Phi^{(B)}(\ell + \frac{1}{2}) = -\Phi^{(A)}(\ell + \frac{1}{2})$. The two sectors thus "sample" these cancelling functions at different energies $n = H_i + \ell$, and it is only by considering the functions themselves — or equivalently by considering $\langle a_{nn}\rangle$ — that the cancellation (2) becomes apparent.

---

*A more precise statement of this result appears in Ref. [1], where a proof is given and certain qualifications are discussed.

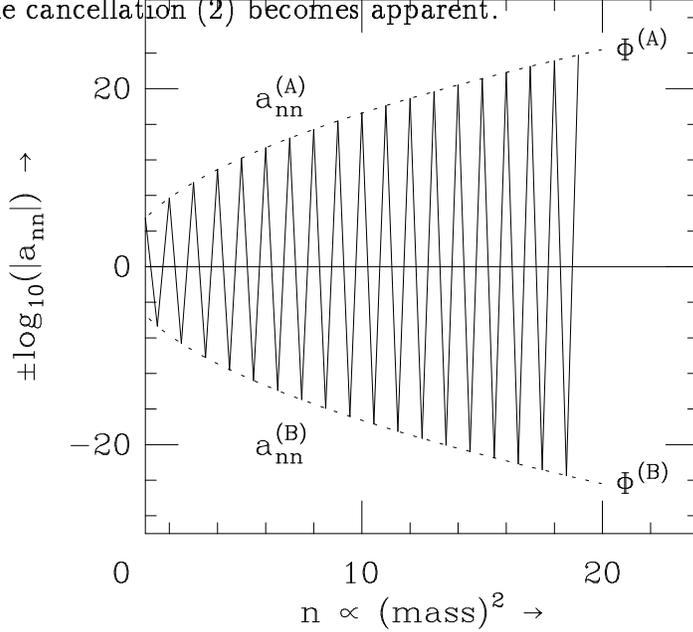

Fig. 1: The net number of physical states $a_{nn}$ for the two-sector model discussed in the text, plotted versus energy $n$ [equivalently the spacetime $(\text{mass})^2$].

In Fig. 1, we have sketched a likely scenario for this toy model, plotting the net physical-state degeneracies $a_{nn}$ as a function of energy $n$. For simplicity, we plot $\pm\log_{10}(|a_{nn}|)$ where the minus sign is chosen if $a_{nn} < 0$ (i.e., if there is a surplus of fermionic states over bosonic states at energy $n$). Although $a_{nn}$ takes values only at the discrete energies $n \in \mathbb{Z}/2$, we have connected these points in order of increasing $n$ to stress the fluctuating oscillatory behavior that $a_{nn}$ experiences as the energy $n$ is increased. Note that as a consequence of the cancellation of the functional forms which describe the separate $A$ and $B$ sectors, the net number of physical states $a_{nn}$ is forced to *oscillate symmetrically around zero* as the energy $n$ increases.

This is a generic consequence of our result, and indeed such oscillatory behavior appears in any modular-invariant tachyon-free theory *regardless* of the number of sectors present. For example, in Fig. 2 we have plotted the number of states for a certain *six*-sector theory whose one-loop partition function appears in Ref. [1]; the vacuum energies of these six sectors $A$ through $F$ are respectively $H_i = 0$, $1/4$, $3/8$, $1/2$, $3/4$ and $7/8$ (modulo 1), and we see that now the six sectors combine to produce a more complicated oscillation pattern. Indeed, from Fig. 2 we see that the numbers of states in the $C$ and $F$ sectors appears to grow with the fastest rate of exponential growth, while the $B$ and $E$ sectors exhibit only moderate exponential growth and the the $A$ and $D$ sectors appear to grow the slowest. The six corresponding functions $\Phi^{(A,...,F)}(n)$ nevertheless have a sum which cancels: the dominant exponential terms within the functions $\Phi^{(C)}$ and $\Phi^{(F)}$ cancel directly; the next-to-dominant exponential terms within $\Phi^{(C)}$ and $\Phi^{(F)}$ combine with the leading exponential terms within $\Phi^{(B)}$ and $\Phi^{(E)}$ to produce a second cancellation; and finally the next-subleading terms which survive these first two cancellations combine with the leading exponential terms within the functions $\Phi^{(A)}$ and $\Phi^{(D)}$ to produce a final cancellation. Thus all of the sectors play a non-trivial role in producing the required cancellation of $\langle a_{nn} \rangle$.

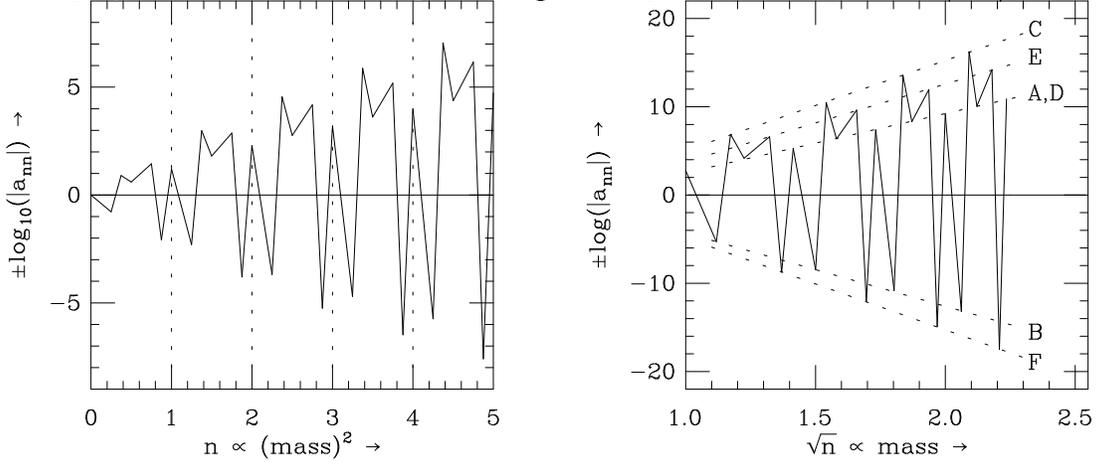

Fig. 2: The net number of states for a theory with six sectors ($A$ through $F$), plotted versus worldsheet energy $n$ and versus spacetime mass $\sqrt{n}$.

It is clear that spacetime supersymmetry appears as a special case of this generic behavior, for in this case we have $a_{nn} = 0$ level-by-level and the "amplitude" of this oscillation is zero. Thus, if spacetime supersymmetry is to be broken in such a way that no physical tachyons are introduced and modular invariance is be maintained (as we would demand in any self-consistent string theory), then our result implies that we can at most "misalign" this bosonic and fermionic cancellation, introducing a mismatch between the bosonic and fermionic state degeneracies at each level in such a way that a "misaligned supersymmetry" survives and the bosonic and fermionic surpluses carefully compensate for each other across the spectrum. It will be interesting to see which classes of physical supersymmetry-breaking scenarios do not lead to such behavior, and are thereby precluded. For example, we can already rule out any scenario in which the energies of bosonic and fermionic states are merely shifted relative to each other by some amount $\Delta n$. Instead, we would need to simultaneously create a certain number $\Phi(n+\Delta n) - \Phi(n)$ of additional states so that the cancellation of the functional forms $\Phi$ is still preserved. Such supersymmetry-breaking scenarios are currently being investigated.

## 3.3  Misaligned Supersymmetry and Finiteness

As discussed in the Introduction, modular invariance and the absence of physical tachyons are the conditions which guarantee finite loop amplitudes in string theory. Since these are also the conditions which yield the "misaligned supersymmetry", it is natural to interpret the resulting boson/fermion oscillation as the mechanism by which the net numbers of states in string theory distribute themselves level-by-level so as to produce finite amplitudes. Let us now briefly provide some evidence for this by focusing on the simplest loop amplitude in string theory, namely the one-loop vacuum polarization amplitude or cosmological constant, defined as

$$\Lambda \equiv \int_{\mathcal{F}} d^2\tau \, (\mathrm{Im}\,\tau)^{-(1+D/2)} \sum_{m,n} a_{mn} \, \overline{q}^m \, q^n \; . \tag{3}$$

Here $D$ is the dimension of spacetime, and $\mathcal{F}$ is the fundamental domain of the modular group, $\mathcal{F} \equiv \{\tau : \ |\mathrm{Re}\,\tau| \leq 1/2, \ \mathrm{Im}\,\tau > 0, \ |\tau| \geq 1\}$.

At first glance, it may seem difficult to relate misaligned supersymmetry to the finiteness of $\Lambda$, for misaligned supersymmetry concerns only the *physical*-state degeneracies $a_{nn}$, and does not involve subsequent integrations over the fundamental domain. Furthermore, we see that Eq. (3) is manifestly finite if the spectrum is tachyon-free (*i.e.*, if $a_{nn} = 0$ for all $n < 0$), since the presumed modular invariance of the theory has already been used to truncate the region of $\tau$-integration to the fundamental domain $\mathcal{F}$ and thereby avoid the dangerous ultraviolet $\tau \to 0$ region. What we require, by contrast, is an expression for $\Lambda$ which depends on only the physical-state degeneracies $a_{nn}$ and for which the finiteness depends on precisely the behavior of these degeneracies.

Fortunately, such an alternative expression exists, and in Ref. [3] it is shown that $\Lambda$ in Eq. (3) may also be rewritten as

$$\Lambda \;=\; \frac{\pi}{3} \lim_{y\to 0} \left\{ y^{1-D/2} \sum_n a_{nn} \, \exp(-4\pi n y) \right\} \; . \tag{4}$$

Here the sum is over all energy levels in the theory, and $y$ serves as a cutoff which regulates the otherwise divergent sum $\sum_n a_{nn}$.

This result (4) is quite powerful, enabling us to formally calculate the complete one-loop cosmological constant (3) given knowledge of only the *physical* state degeneracies. More importantly, however, Eq. (4) provides a severe constraint on the distributions of the physical states in any tachyon-free modular-invariant theory, for somehow the net degeneracies $\{a_{nn}\}$ must arrange themselves in such a way that the right side of Eq. (4) is finite. In particular, for $D > 2$, the finiteness of $\Lambda$ requires that the $a_{nn}$ satisfy the constraint

$$\lim_{y \to 0} \sum_n a_{nn} \exp(-4\pi n y) = 0 \ . \tag{5}$$

*A priori*, there are any number of conceivable distributions $\{a_{nn}\}$ which might satisfy Eq. (5), and therefore this condition alone is not sufficiently restrictive to *predict* the resulting behavior for the actual distribution of physical states. The condition (5) can nevertheless be used to *rule out* certain behavior for the net degeneracies $a_{nn}$. For example, it is straightforward to show that if $a_{nn} \sim n^{-B} e^{C\sqrt{n}}$ with $C > 0$ as $n \to \infty$, then the left side of Eq. (5) diverges as $y^{2B-3/2} \exp(C^2/16\pi y)$ as $y \to 0$. Thus all direct exponential growth for the net degeneracies $\{a_{nn}\}$ is prohibited. Similar conclusions also hold for polynomially growing $a_{nn}$.

Without knowledge of misaligned supersymmetry and the resulting boson/fermion oscillation, these last results would seem quite remarkable, since we know that each individual sector contributes a set of degeneracies $\{a_{nn}^{(i)}\}$ which *does* grow exponentially. Indeed, it is precisely this exponential growth which is responsible for the famous Hagedorn phenomenon which is thought to signal a phase transition in string theory.

However, misaligned supersymmetry and its implicit boson/fermion oscillation now provide a natural alternative solution which reconciles the finiteness condition (5) with the growing behavior for which $|a_{nn}| \to \infty$ as $n \to \infty$. Indeed, we can easily see that with such oscillations, *many* growing solutions to Eq. (5) are now possible. For example, simple distributions such as $a_{nn} = (-1)^n n^2$, $a_{nn} = (-1)^n n^4$, $a_{nn} = (-1)^n (n - n^5)$, and $a_{nn} = (-1)^n (2n^3 + n^5)$ all non-trivially satisfy Eq. (5), where the factor of $(-1)^n$ is meant to illustrate the alternating-sign behavior for $a_{nn}$ which is characteristic of the boson/fermion oscillations. In fact, Ref. [1] contains a detailed analysis of the behavior of a certain set of modular-invariant, exponentially growing, oscillating degeneracies $\{a_{nn}\}$ for which the corresponding one-loop cosmological constant *vanishes exactly* [4]. Thus, we see that it is precisely the misaligned supersymmetry and its carefully balanced boson/fermion oscillation which enable solutions with $|a_{nn}| \to \infty$ as $n \to \infty$ to satisfy Eq. (5), and which serve the crucial mechanism by which finiteness is actually maintained level-by-level in the string spectrum.

## 4   Conclusions and Applications

We have provided a non-technical overview of misaligned supersymmetry [1], a powerful constraint on the numbers and distribution of physical states in string theory. We demonstrated that this phenomenon naturally constrains the degree to which spacetime supersymmetry may be broken in string theory, and showed that the existence of misaligned supersymmetry is necessary for the finiteness of string amplitudes. Furthermore, misaligned supersymmetry is completely generic, and appears in *any* theory which is modular-invariant and free of physical tachyons. It thus applies to superstrings, to heterotic strings compactified via *any* mechanism to *any* spacetime dimension, or even to strings with unusual worldsheet conformal field theories (such as fractional superstrings). This is therefore a general "stringy" phenomenon with many applications. For example, misaligned supersymmetry should be particularly relevant to any system in which the asymptotic number $\langle a_{nn} \rangle$ of high-energy states plays a crucial role, such as in string thermodynamics and the possible string phase transition. There may even be applications to *hadron-scale* string theory [5].

There are nevertheless a number of potential extensions to our results. For example, we would like to understand the role that misaligned supersymmetry plays in ensuring finiteness to *all* orders (not just one-loop), and also for all $n$-point functions. Clearly, this requires extending our results to include the *unphysical* string states, as well as string *interactions*. Also, we would also like to use misaligned supersymmetry to formulate a more precise constraint on physical supersymmetry-breaking scenarios: precisely which classes of dynamical mechanisms can yield the required boson/fermion oscillations? Clearly this will require an understanding of misaligned supersymmetry as a *symmetry*, rather than merely as a constraint on the numbers of states. Such work is in progress.

## Acknowledgments

I am pleased to thank A. Anderson, D. Kutasov, C.S. Lam, R. Myers, D. Spector, H. Tye, and E. Witten for discussions on the work described here. Portions of this research were performed while the author was at McGill University. This research was supported in part by NSERC (Canada), FCAR (Québec), and DOE Grant No. DE-FG-0290ER40542 (US).